\documentclass[aps,pra,twocolumn,groupedaddress]{revtex4}
\usepackage{graphicx}

\begin{document}

\title{Simple, broadband, optical spatial cloaking of very large objects}


\author{John C. Howell$^1$}
\author{J. Benjamin Howell$^2$}
%\email[]{howell@pas.rochester.edu}
%\thanks{}
\affiliation{$^1$Department of Physics and Astronomy, University of Rochester, Rochester, New York 14627, USA}
\affiliation{$^2$TCMS, Rochester, New York 14618, USA}


%\date{\today}

\begin{abstract}
We demonstrate three simple cloaking devices that can hide very large spatial objects over the entire visible spectrum using only passive, off-the-shelf optics.  The cloaked region for all of the devices exceeds $10^6$ mm$^3$ with the largest exceeding $10^8$ mm$^3$.  Although uni-directional, these cloaks can hide the cloaked object, even if the object is transversely or self-illuminated.  Owing to the small usable solid angle, but simple scaling, these cloaks may be of value in hiding small field-of-view objects such as mid- to high-earth orbit satellites.
\end{abstract}


\maketitle 

\section{Introduction}
The intriguing and exciting possibilities of optical spatial cloaking have attracted both the popular culture and the scientific community \cite{Pendry06,Leonhardt06,Chen10,Schurig06,Cai07,Liu08,Smolyaninov08,Smolyaninov09,Landy12,Liu09,Valentine09,Gabrielli09,Leonhardt09,Zhang11}.  Great strides have been taken to achieve the ``Holy Grail" of optical cloaking: broadband optical invisibility, omni-directionality, the ability to cloak macroscopic objects and be invisible itself. Much of the initial developments in experimental optical cloaking were based on transformation optics \cite{Pendry06,Leonhardt06,Chen10} and studied in metamaterials \cite{Pendry06,Schurig06,Cai07,Liu08,Liu09,Landy12,Smolyaninov08,Smolyaninov09} and in patterned dielectrics \cite{Valentine09,Gabrielli09} with the use of quasi-conformal mapping.  While remarkable progress has been made, much work remains in terms of achieving the cloaking at optical wavelengths, increasing the bandwidth over which the cloak works, and scaling to large dimensions. Towards these goals, a recent classical optics method using birefringent calcite crystals \cite{Zhang11} achieved a polarization-dependent, broadband, visible, uni-directional cloak of a small incline with a peak height of 2mm.  Another intriguing advance was the recent demonstration of temporal cloaking (hiding a temporal event) work \cite{Fridman12} based on split temporal lensing and dispersive propagation in fibers.

Here, we report on cloaking devices based on off-the-shelf optics, which can achieve all of the elements of optical spatial cloaking except omni-directionality.  We demonstrate cloaking over the visible spectrum with cloaking regions exceeding $10^6$ mm$^3$ with the largest exceeding $10^8$ mm$^3$.  At its most basic level, a cloaking device simply guides light around an object as if the object isn't there.  One might argue that an endoscope, an index-guiding fiber system used to image hollow organs in the human body, achieves this end.  The etymology of ``smoke and mirrors" implies the use of distraction and optical illusion through the careful guiding of light.  Based on this history with optical manipulations and illusions \cite{Houdin1868}, it may not come as a surprise that simple passive optics can be used to cloak very large optics.  To demonstrate this point, we built and report on three cloaking devices. Importantly, each device can be easily scaled to much larger systems.

The first device is based on the bending of light at a dielectric interface.  We demonstrate a simple realization of the device using two L-shaped water-filled tanks.  The second device uses quadratic phase elements, such as lenses, and doesn't suffer from the edge effect problems of the first device (this is the spatial equivalent of the temporal cloak \cite{Fridman12}).  The third device simply uses mirrors to guide light around the object.  The latter cloak, while not new \cite{Houdin1868,MirrorInvisibility}, is to demonstrate the ease of scaling of these systems.

\section{Cloaking with Snell's law}

\begin{figure}[h]
 \begin{centering}
  \includegraphics[scale=0.35]{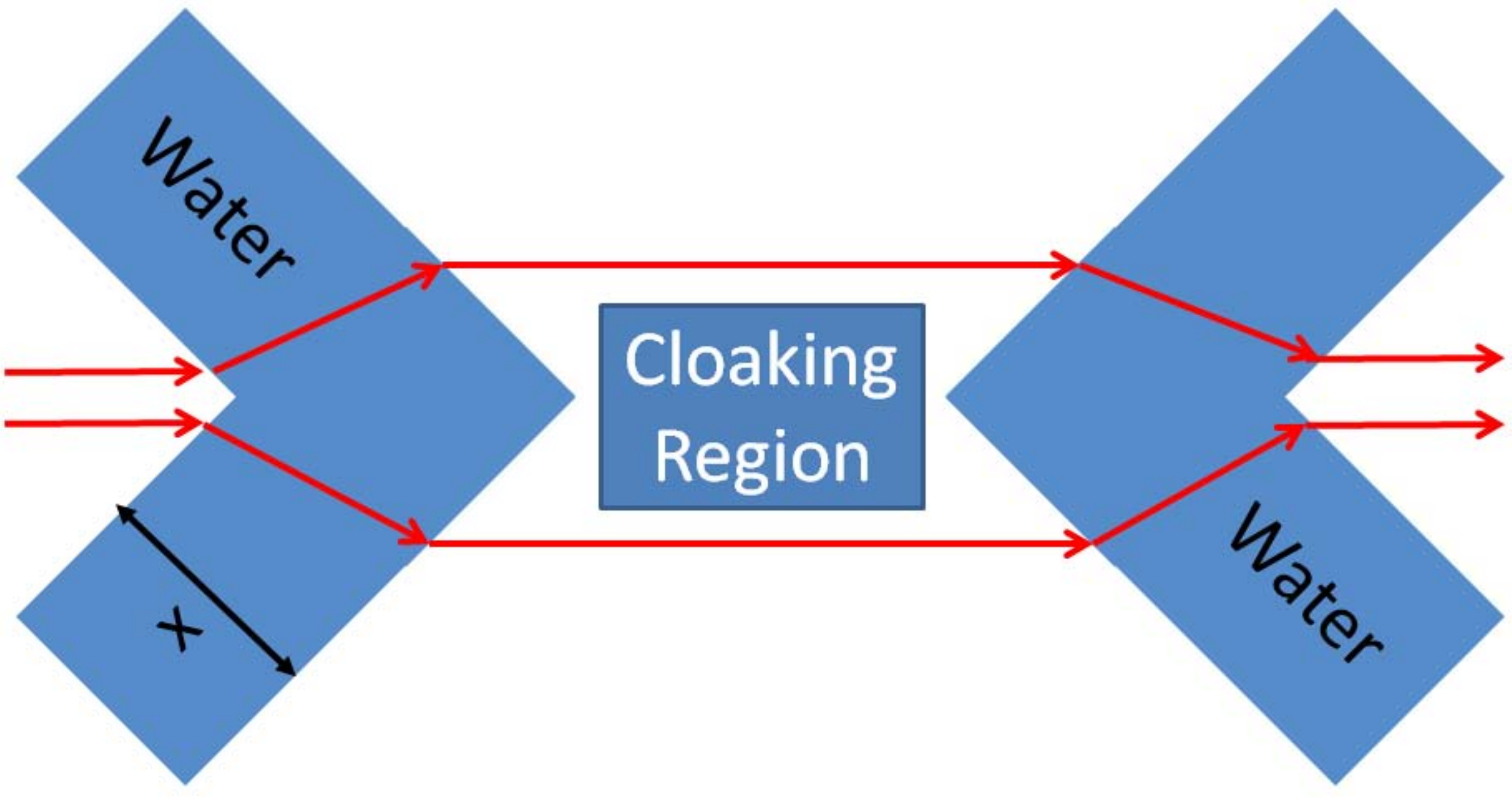}
  \caption{A cloaking device based on Snell's law. Two L-shaped, water-filled tanks bend light around a cloaking region and bring the light back with the same direction and displacement.  At the first interface, light bends away from the optical axis due to the interface with water.  The light then bends back in the same direction at the second interface.  For rays parallel to the optical axis, the second tank brings the light back with both the same direction and no displacement.}\label{fig:typeI}
  \end{centering}
\end{figure}

\begin{figure}[h]
 \begin{centering}
  \includegraphics[scale=0.4]{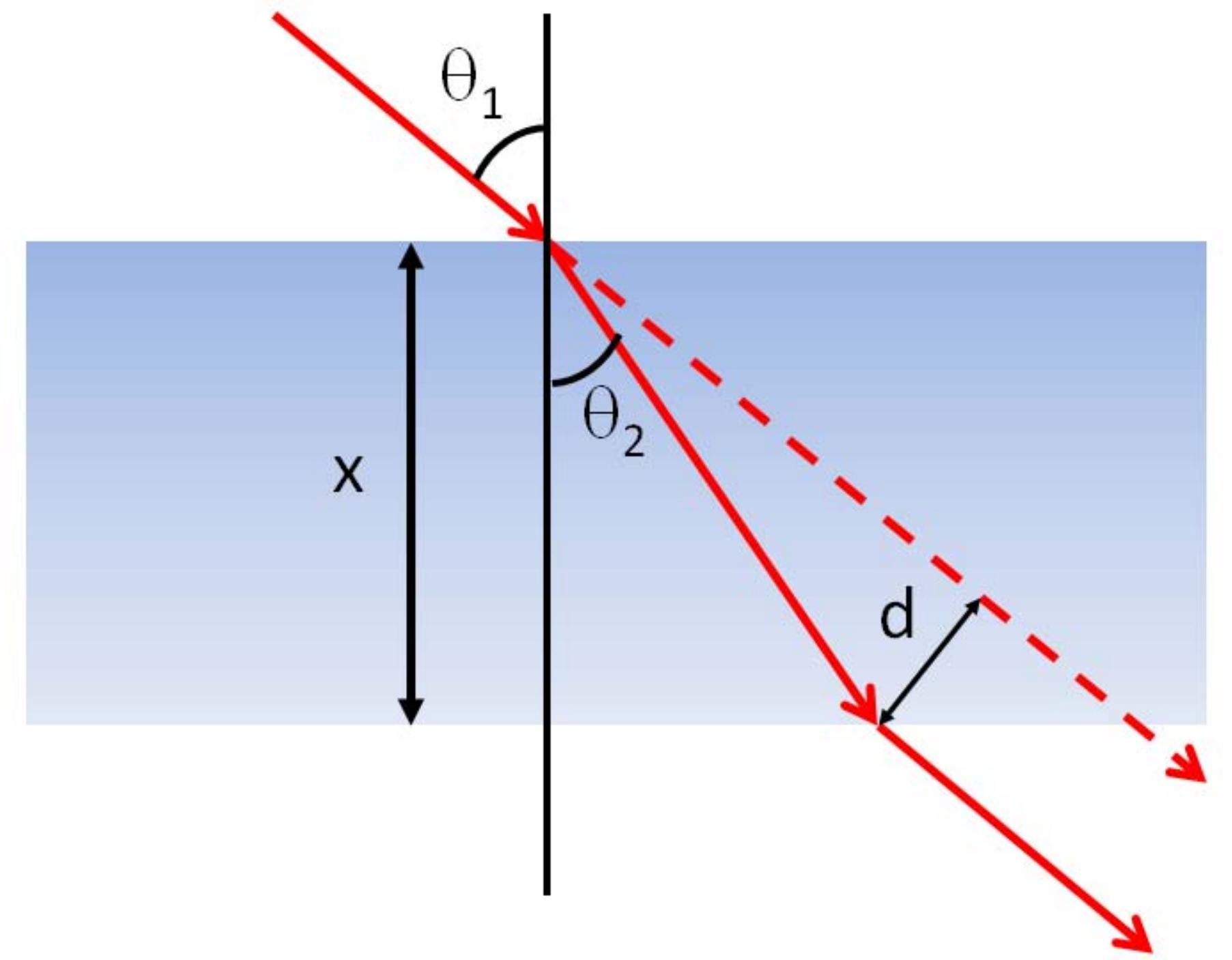}
  \caption{The distance $d$ between an unaltered ray and the path of the ray after exiting the water can be found from Snell's law and the width of the tank $L$.}\label{fig:snellslaw}
  \end{centering}
\end{figure}

The experimental setup of the first device is shown in Fig. \ref{fig:typeI}.  The cloak is based on the idea that light is transversely shifted, but has the same direction, after passing through a tilted medium (having different index of refraction) with two straight parallel faces. The setup uses two water-filled tanks to bend light around a cloaking region.  The first tank causes all rays below the optical axis to be shifted downward and the rays above the optical axis to be shifted upwards.  After the rays exit the second wall of the first tank, they are parallel to the optical axis, but have a displacement relative to their original trajectory.  The displacement hides any object in the middle.  The second tank undoes the effects of the first tank so that the rays are once again parallel with the displacement undone.  The diameter of the cloaking region can be found from Snell's law and the width of the tank.  From Fig. \ref{fig:snellslaw}, we can compute the size of the cloaking region.  We use Snell's law $n_{a}\sin(\theta_a)=n_w \sin(\theta_w)$, where $n_a=1$ ($n_w=1.33$) is the index of refraction of air (water) and $\theta_a$ ($\theta_w$) is the angle of the ray relative to the normal in air (water).  Using straightforward trigonometry,  we find that the distance $d$ between the incident ray and the ray after propagating through the medium is given by
\begin{equation}
d=x \frac{\sin(\theta_a-\arcsin(n_w^{-1}\sin(\theta_a)))}{\cos(\arcsin(n_w^{-1}\sin(\theta_a)))}.
\end{equation}
The tanks have a width of $x\approx$200 mm and the incident light rays have an angle of $\pi/4$ yielding a cloaking region of $2d=105$mm in good agreement with the observed size.

An alternative perspective to the functionality of this cloak is that the effect of the L-shaped tanks is to create a split linear phase ramp across the front of the electric field.  The split linear phase ramp causes parallel rays to be deflected in opposite directions about the cloaking region.  This is a uni-directional cloaking device because some rays not parallel to the optical axis can penetrate the cloaking region.  In addition, the non-parallel rays traverse different lengths of water and have different incident angles in the two tanks causing distortion to the background.  Further, unlike other cloaks and the second device discussed later, there isn't an effective compression of the field.  This lack of field compression leads to edge effects.  This first device will suffer from edge effects at the extreme wings of the device. Owing to the deflection without compression of the rays at the first interface, the second interface of the tank has to protrude further than the first interface.  So, unless the tanks are infinite in extent, the edges will have some problems.  It should also be noted that transverse- or self-illumination of the cloaked object cannot be observed in the cloaked direction since the tanks will deflect the light away from the observer.

\begin{figure}[h]
 \begin{centering}
   \includegraphics[scale=0.3]{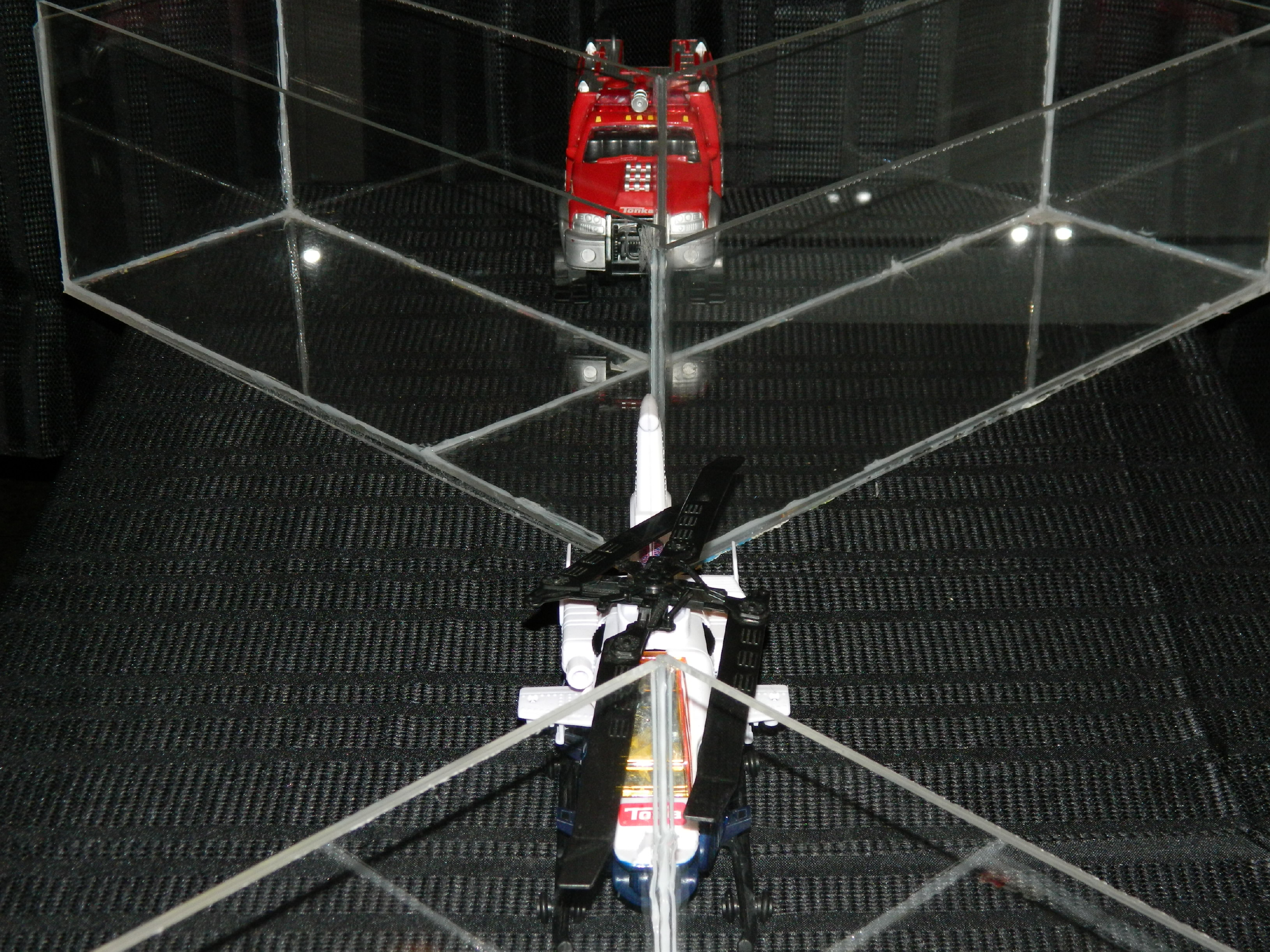}
  \caption{An aerial view of the first cloaking device.  A helicopter is shown inside the cloaking region of the first device.  A truck is shown on the other side of the viewing region.  The truck will appear in the helicopters place when water is inside the tanks and viewed along the optical axis. }\label{cloakedhelisetup}
  \end{centering}
\end{figure}

\begin{figure}[h]
 \begin{centering}
   \includegraphics[scale=0.33]{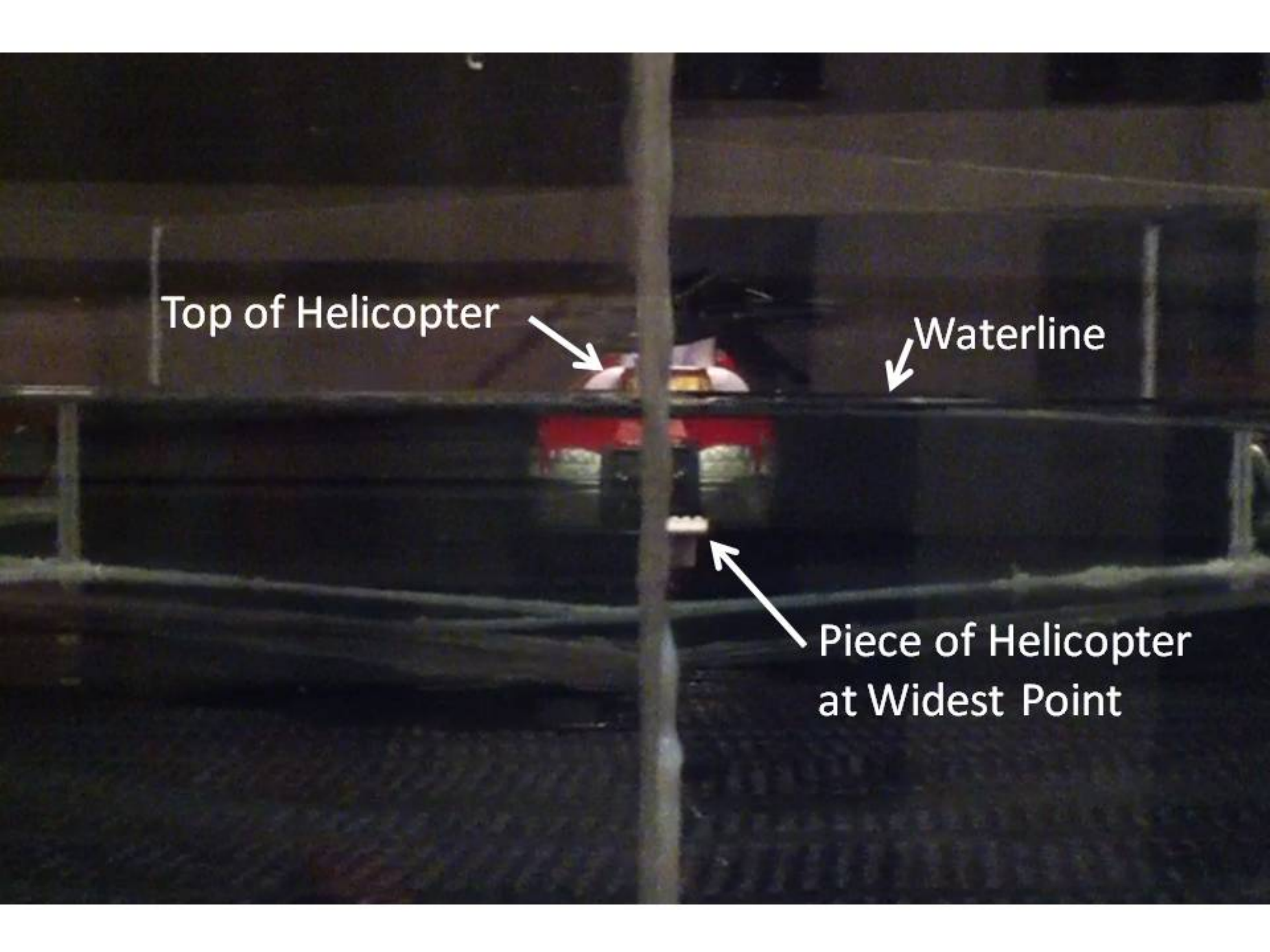}
  \caption{Below the waterline, the helicopter is cloaked and the truck appears in its place.  The light coming from the truck passes around the helicopter, via the water tanks, and is then seen in the place of the helicopter. Above the waterline, the helicopter is shown in front of the truck. }\label{cloakedheli}
  \end{centering}
\end{figure}

The actual experimental setup is shown in Fig. \ref{cloakedhelisetup}.  An aerial view of the setup shows a helicopter in the cloaking region between the two water tanks.  The cloaking ability of the first cloaking device is shown in Fig. \ref{cloakedheli}.  Below the waterline in the picture, the helicopter cannot be seen and the truck appears in its place.   However, above the waterline, the helicopter is visible and in front of the truck.  The bending of the light causes the light to pass around the bottom of the helicopter and bend back so that the truck can be viewed in its place.  The transverse width of the helicopter is approximately 125 mm, being approximately 20 mm larger than the cloaking diameter.  A small, white, brightly lit piece of the helicopter at its widest point can be seen.  It should be noted that the helicopter is illuminated directly from above and remains invisible.  Thus, the transverse illumination of the cloaked object does not reveal its presence, as one would expect from a cloaking device.

While the first cloaking device is straightforward to implement, the scaling to very large objects becomes rather impractical unless one wishes to carry around very large tanks of water.  A generalization of this device is to have a device which causes a split linear phase ramp.  This can be achieved with a hologram, a spatial light modulator, a large piece of glass with long prism-like wedges (equivalent to a Fresnel lens but with linear rather than quadratic etching).

\begin{figure}[h]
  \begin{centering}
  \includegraphics[scale=0.45]{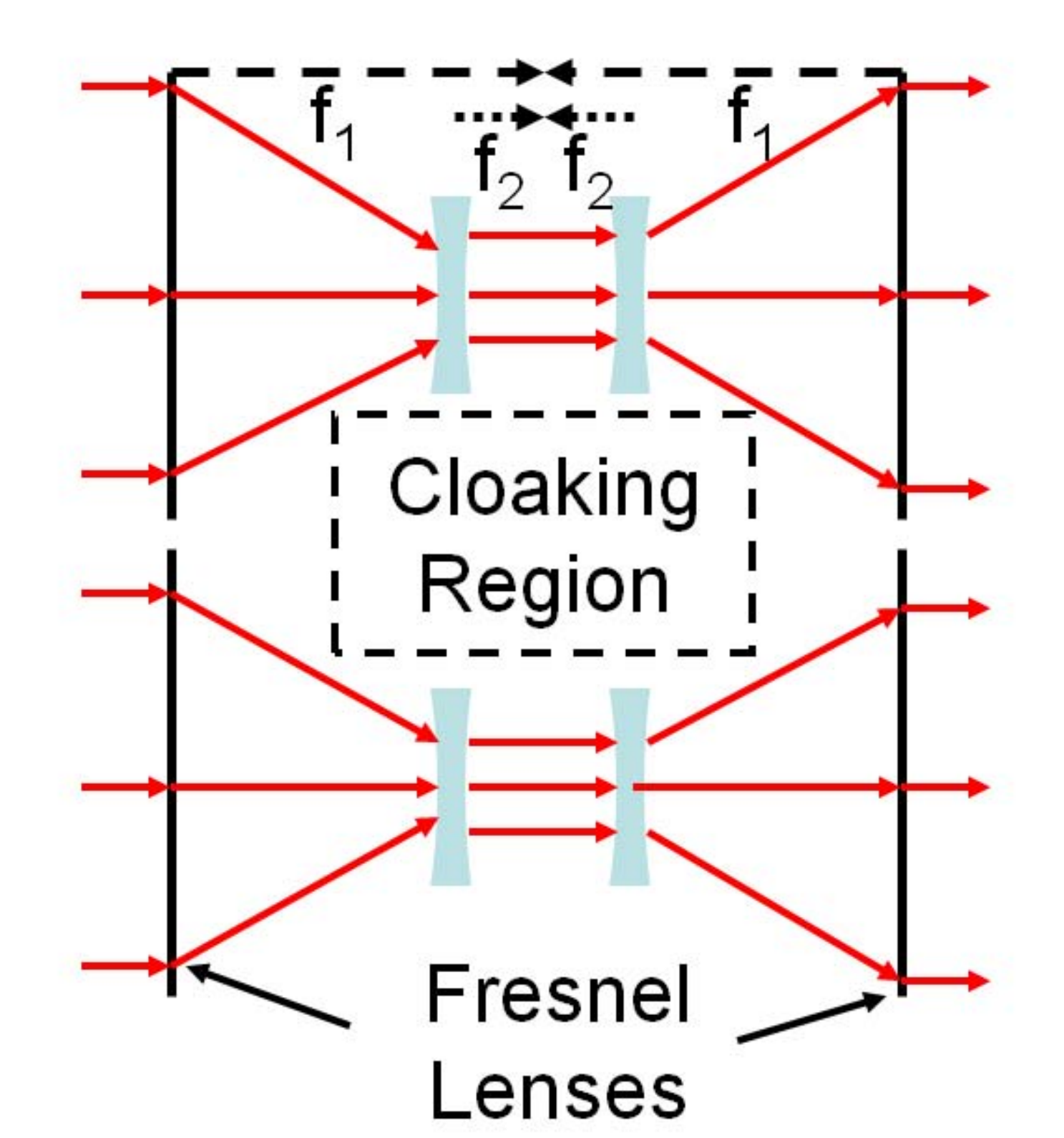}
  \caption{Experimental cloaking schematic of the second device.  Converging and diverging lenses are used to map light around the cloaking region.}\label{fig:typeIIexperiment}
  \end{centering}
\end{figure}

\section{Cloaking with lenses}
The second scheme is shown in Fig. \ref{fig:typeIIexperiment}.  This scheme can be considered as the spatial equivalent of the temporal cloak used in \cite{Fridman12}.  Lenses are used to guide light around the cloaking region.  Fresnel lenses are used at the interfaces of the cloaking device, because of their low mass, scalability and their rectangular shape.   Unfortunately, passing through a focus inverts the background behind the object.  In this spatial cloak, diverging lenses between the Fresnel lenses prevent the light from passing through a focus, which means that the image will be upright rather than inverted.  They also have the interesting property that the separation between lenses can be quite large after the rays are collimated, so that the cloaking region can be extended longitudinally.

\begin{figure}[h]
  \begin{centering}
  \includegraphics[scale=0.35]{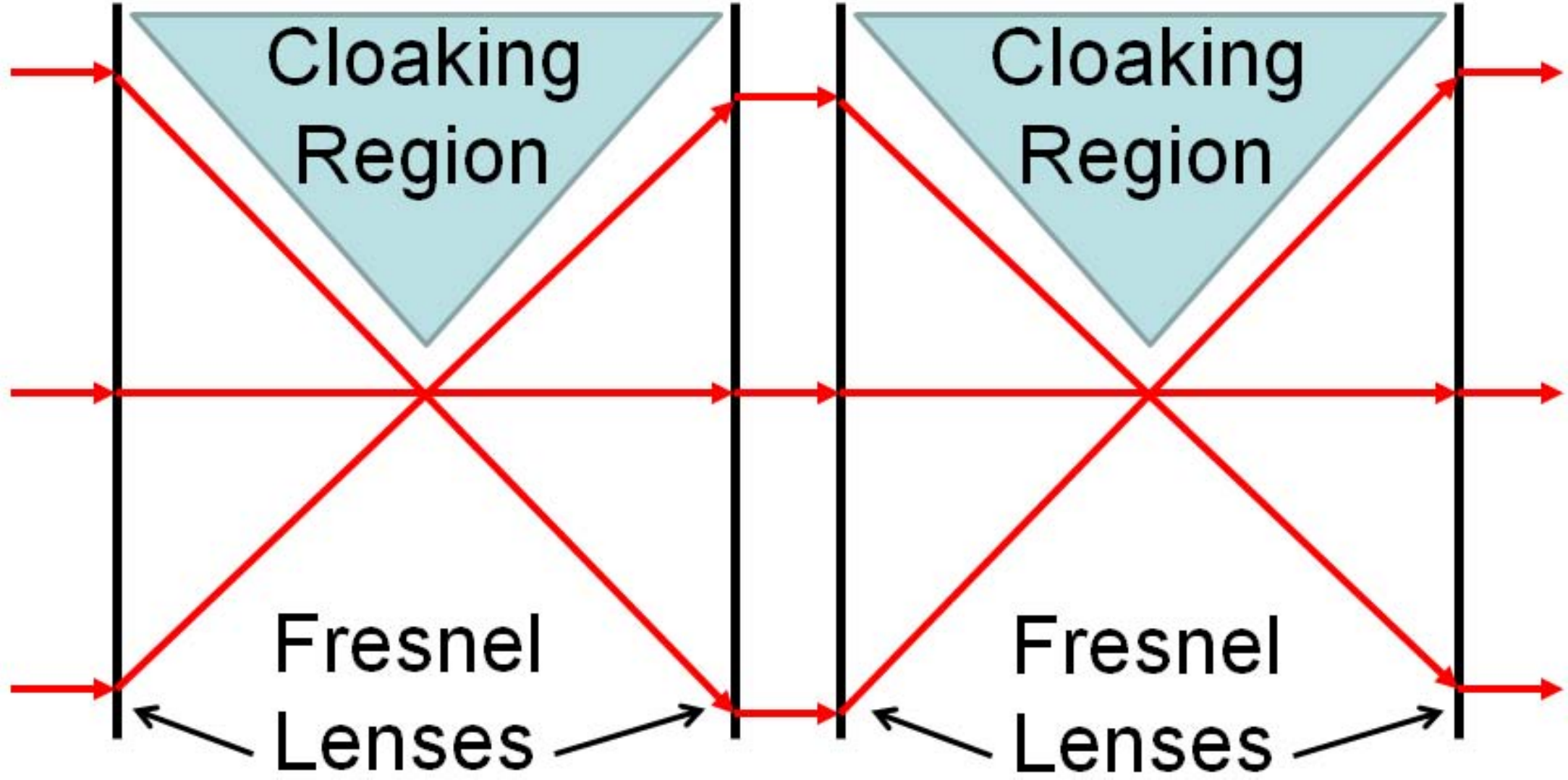}
  \caption{The alternative schematic for the second device. Four Fresnel lenses in series are used.  Two sets of Fresnel lenses with each set separated by twice the focal length make it so the image is not inverted.  The distance between lens pairs can be arbitrarily small.  In the actual experiment, they were mounted together as if it were a single lens. }\label{type2alternative}
  \end{centering}
\end{figure}

\begin{figure}[h]
  \begin{centering}
  \includegraphics[scale=0.55]{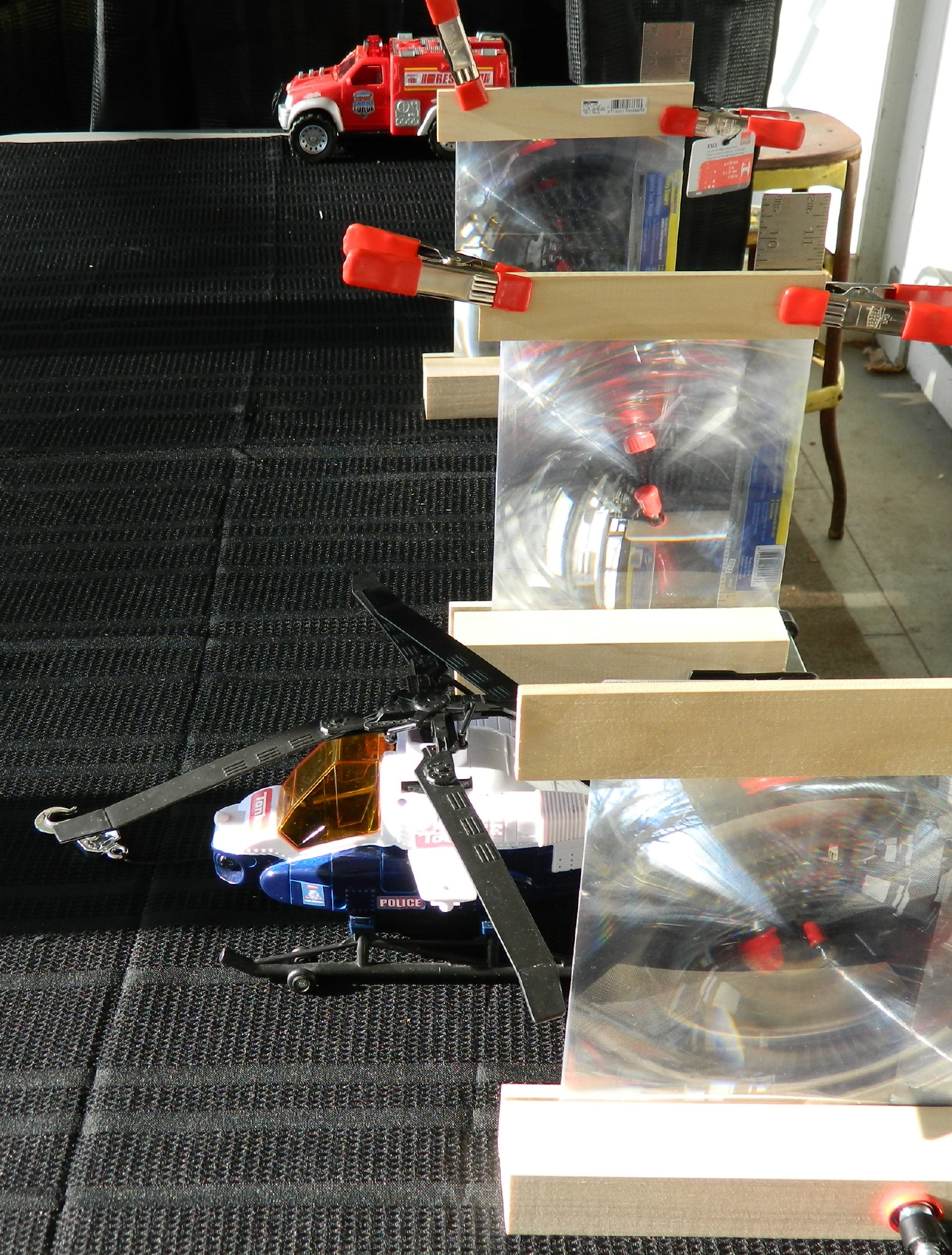}
  \caption{The setup for the second cloaking device is shown.  The tail of a helicopter is at the focus of a Fresnel lens (light passes around it).  Four Fresnel lenses (the two in the middle are in contact) allow for a 1 to 1 noninverted imaging of the background.  The lenses have the dimensions of 175mm x250mm and have a focal length of 200mm..  The truck is placed a distance of 750mm from the last lens. }\label{type2setup}
  \end{centering}
\end{figure}

\begin{figure}[h]
  \begin{centering}
  \includegraphics[scale=0.35]{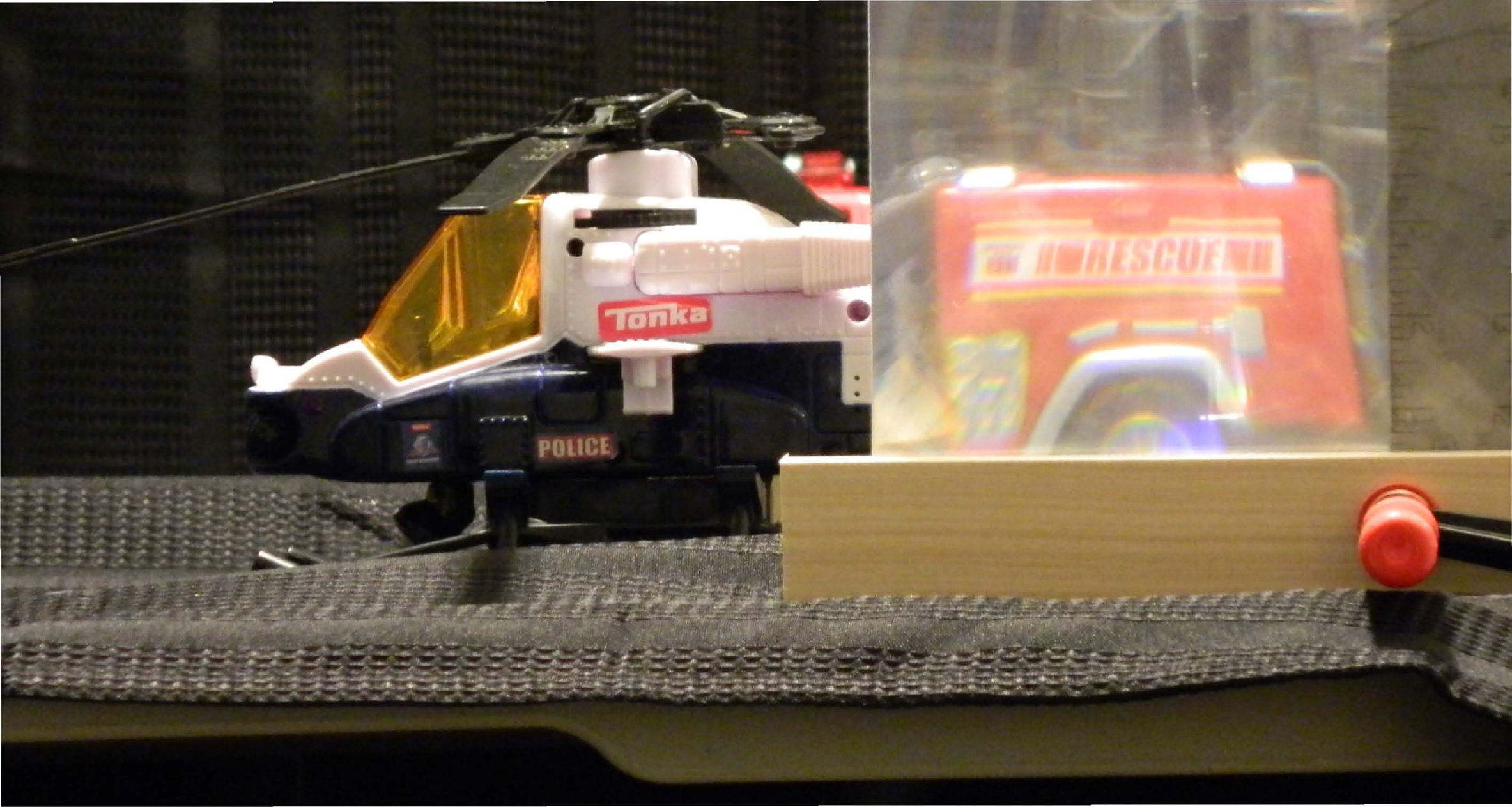}
  \caption{Viewing along the optical axis, we see the truck appearing in the place of the tail of the helicopter. The observation was done with a 21x magnification camera at 6.4m from the first lens.}\label{cloakedhelitype2}
  \end{centering}
\end{figure}

An alternative, but slightly more bulky design was actually used to demonstrate the cloaking. The alternative design, as shown in Fig. \ref{type2alternative}, removes the two diverging lenses from the setup.  However, with the two diverging lenses removed the image of the background is inverted.  To make the image upright, we use another set of Fresnel lenses also separated by 2$f_1$ in series with the first two lenses. In the actual experiment, the two middle Fresnel lenses were mounted together as if they were a single lens.  The design of the device is shown in Fig. \ref{type2setup} and cloaked helicopter is shown in Fig.\ref{cloakedhelitype2}. It can be seen that the tail of the helicopter is cloaked and the truck, behind the helicopter, appears in its place.  One can notice in the uncloaked region, a small portion of the truck in the background can be seen above the helicopter.  Four 175mm x250mm Fresnel lenses were used, each having a focal length of 200mm.  The truck is placed a distance of 750mm from the first lens and observation is done with a 21x magnification camera at $\approx$6.4m from the rear lens.  The image quality is limited by the quality of the Fresnel lenses.

\section{Cloaking with mirrors}
\begin{figure}[h]
  \begin{centering}
  \includegraphics[scale=0.45]{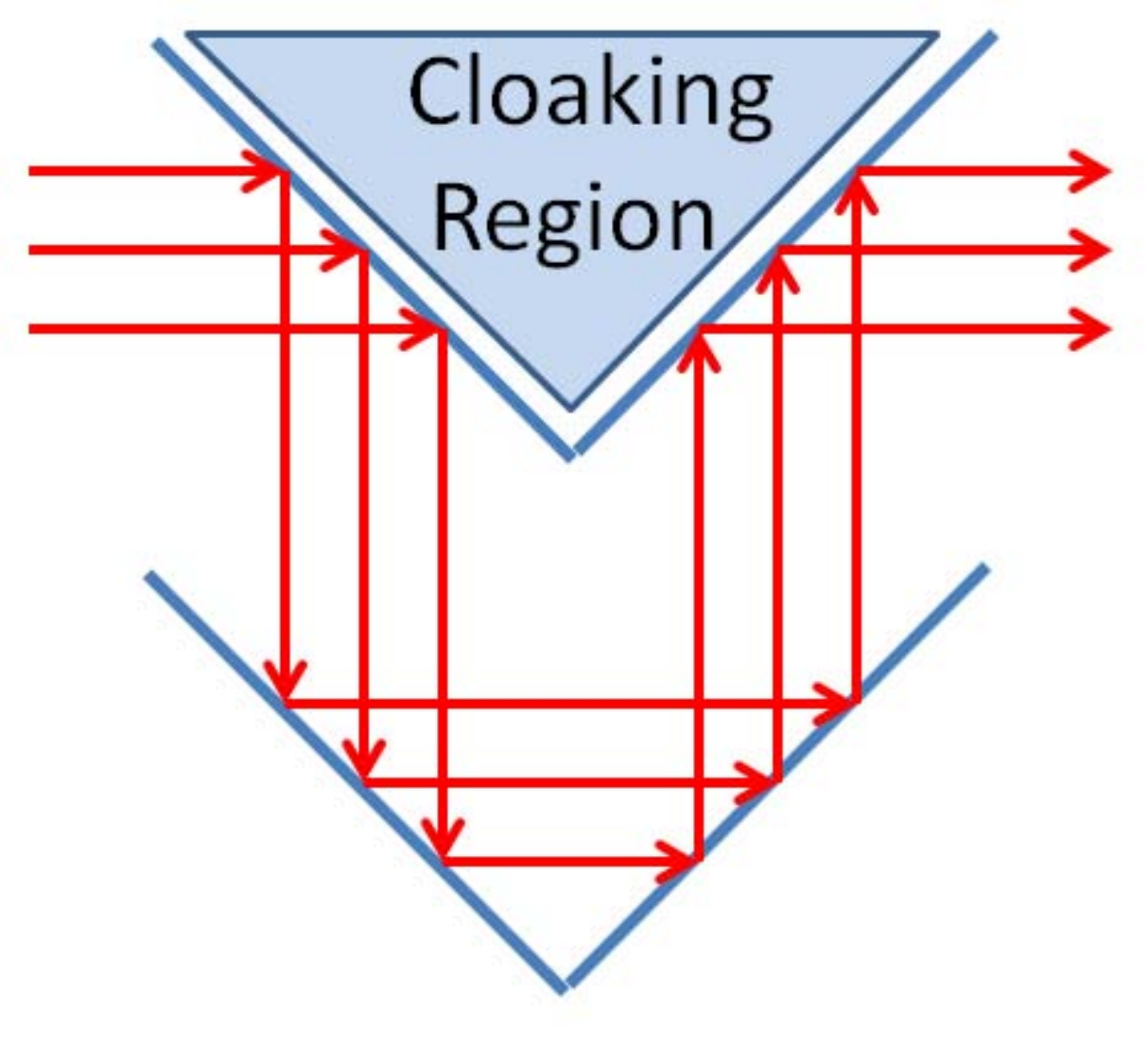}
  \caption{The schematic for the third device is shown.  Mirrors reflect light around the cloaking region.}\label{device3schematic}
  \end{centering}
\end{figure}

The last cloak is probably the most obvious design one would use to make light pass around an object.  The design of the device is shown in Fig. \ref{device3schematic}.  Invisibility with mirrors has been done and are youtube hits \cite{MirrorInvisibility}. The point we wish to emphasize is not the novelty but the ease of scaling to nearly arbitrary size.  The first mirror reflects the light away from the cloaking region.  The light then bounces off of a retro-reflecting mirror pair (two mirrors at right angles).  The light then reflects off of the mirror behind the object.  The retro-reflecting mirrors make it so that rays leaving the cloak, even off-axis rays, leave at the same angle, albeit with transverse shifts.  The transverse shifts mean that this device optimally works at infinity (large distance from the observer). However, there are some downsides to this cloak. First, there are edge effects if one moves to the side (uni-directional).  Second, the retro-reflecting mirror pair make the cloak visible.  Third, while the rays leave in the same direction, they leave with a displacement that is proportional to the incoming angle relative to the optical axis.  Although having some drawbacks, it is clearly scalable to very large dimensions.

\begin{figure}[h]
  \begin{centering}
  \includegraphics[scale=0.35]{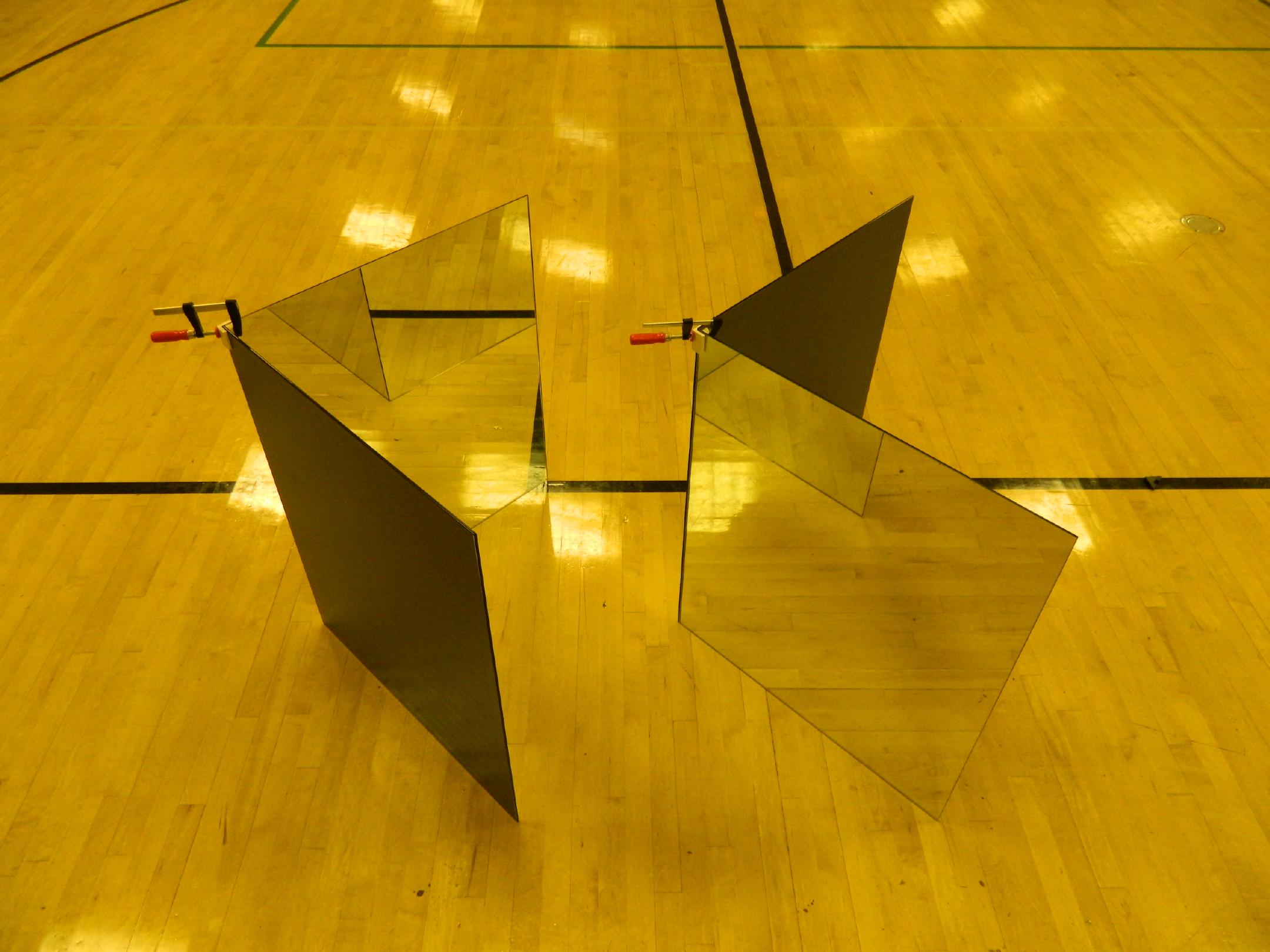}
  \caption{ The setup for the third device is shown. Two sets of right angle mirrors guide light around the cloaked region.}\label{device3setup}
  \end{centering}
\end{figure}

\begin{figure}[h]
  \begin{centering}
  \includegraphics[scale=0.35]{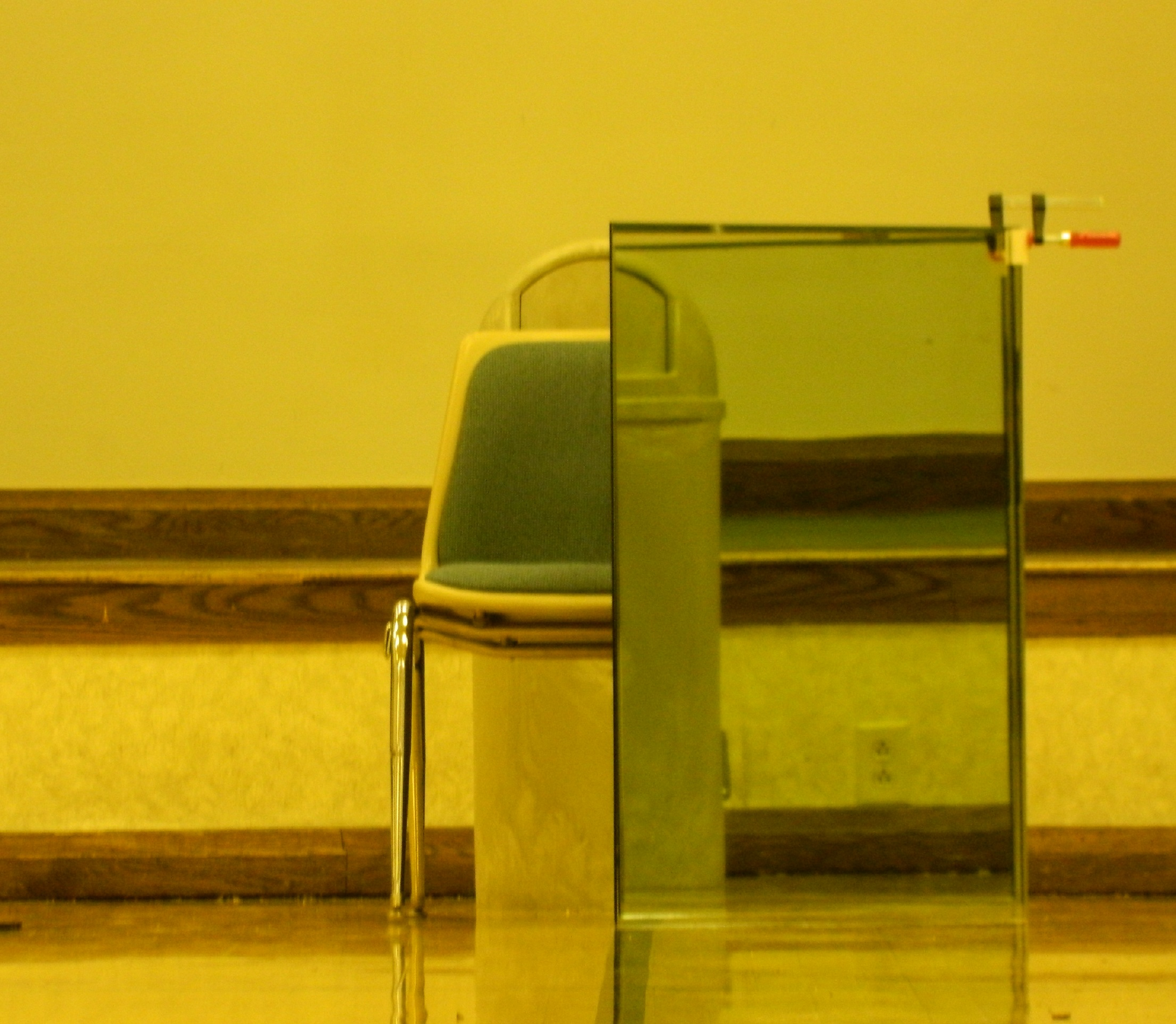}
  \caption{A chair is cloaked and a rubbish can appears in its place.  }\label{cloakedchairtype3}
  \end{centering}
\end{figure}

The actual experiment is shown in Fig. \ref{device3setup}.  Two sets of mirrors are joined at right angles.  It is important to align the mirrors so that the front and rear mirrors are perpendicular.  To minimize background distortion, careful effort was given to securing the joined mirrors at right angles and to be placed as vertical as possible. In Fig. \ref{cloakedchairtype3} one can see in the figure that part of the chair is cloaked and the rubbish can in the background appears in its place.   The image was taken approximately 25 meters from the mirrors, thus ensuring that the cloak occupied a small field-of-view (uni-directionality).  The mirrors have dimensions of 600 mm by 900 mm having a total cloaking volume of $1.6\times 10^8$ mm$^3$.  This volume is sufficient to cloak a human, albeit with not as much convenience as Harry Potter's cloak.  The simplicity of the device means that much larger cloaking devices can easily be built.

\section{Conclusion}
In summary, we have demonstrated three straightforward optical cloaking devices.  The cloaks work over the visible spectrum, have cloaking regions exceeding $10^6$ mm$^3$ and with good coatings, the second device can be made nearly invisible.  The second device does not suffer from edge effects for straight-on viewing.  The downside is that all of these devices are only uni-directional.    The devices may have value, for example, in cloaking satellites in mid- to high-earth orbit or for any low field-of-view cloaking.  It should be pointed out that transverse- or self-illumination of the cloaked object still renders the object invisible to the observer.  While it has been shown that perfect invisibility cannot be attained \cite{Wolf93}, an open question is if standard optics can achieve geometric (ray optic) omni- or multi-directional cloaking \cite{Leonhardt06}.  The authors believe that a cloak with spherical symmetry (much like retro-reflecting spheres achieve multi-directional reflection) may achieve this end.

The authors would like to thank C. Levit for pointing out reference \cite{Houdin1868}.

This work was supported by the HPL.

\bibliography{Cloakbib}

\begin{thebibliography}{18}
\expandafter\ifx\csname natexlab\endcsname\relax\def\natexlab#1{#1}\fi
\expandafter\ifx\csname bibnamefont\endcsname\relax
  \def\bibnamefont#1{#1}\fi
\expandafter\ifx\csname bibfnamefont\endcsname\relax
  \def\bibfnamefont#1{#1}\fi
\expandafter\ifx\csname citenamefont\endcsname\relax
  \def\citenamefont#1{#1}\fi
\expandafter\ifx\csname url\endcsname\relax
  \def\url#1{\texttt{#1}}\fi
\expandafter\ifx\csname urlprefix\endcsname\relax\def\urlprefix{URL }\fi
\providecommand{\bibinfo}[2]{#2}
\providecommand{\eprint}[2][]{\url{#2}}

\bibitem[{\citenamefont{Pendry et~al.}(2006)\citenamefont{Pendry, Schurig, and
  Smith}}]{Pendry06}
\bibinfo{author}{\bibfnamefont{J.~B.} \bibnamefont{Pendry}},
  \bibinfo{author}{\bibfnamefont{D.}~\bibnamefont{Schurig}}, \bibnamefont{and}
  \bibinfo{author}{\bibfnamefont{D.~R.} \bibnamefont{Smith}},
  \bibinfo{journal}{Science} \textbf{\bibinfo{volume}{312}},
  \bibinfo{pages}{1780} (\bibinfo{year}{2006}).

\bibitem[{\citenamefont{Leonhardt}(2006)}]{Leonhardt06}
\bibinfo{author}{\bibfnamefont{U.}~\bibnamefont{Leonhardt}},
  \bibinfo{journal}{Science} \textbf{\bibinfo{volume}{312}},
  \bibinfo{pages}{1777} (\bibinfo{year}{2006}).

\bibitem[{\citenamefont{Chen et~al.}(2010)\citenamefont{Chen, Chan, and
  Sheng}}]{Chen10}
\bibinfo{author}{\bibfnamefont{H.}~\bibnamefont{Chen}},
  \bibinfo{author}{\bibfnamefont{C.~T.} \bibnamefont{Chan}}, \bibnamefont{and}
  \bibinfo{author}{\bibfnamefont{P.}~\bibnamefont{Sheng}},
  \bibinfo{journal}{Nat Mater} \textbf{\bibinfo{volume}{9}},
  \bibinfo{pages}{387} (\bibinfo{year}{2010}).

\bibitem[{\citenamefont{Schurig et~al.}(2006)\citenamefont{Schurig, Mock,
  Justice, Cummer, Pendry, Starr, and Smith}}]{Schurig06}
\bibinfo{author}{\bibfnamefont{D.}~\bibnamefont{Schurig}},
  \bibinfo{author}{\bibfnamefont{J.~J.} \bibnamefont{Mock}},
  \bibinfo{author}{\bibfnamefont{B.~J.} \bibnamefont{Justice}},
  \bibinfo{author}{\bibfnamefont{S.~A.} \bibnamefont{Cummer}},
  \bibinfo{author}{\bibfnamefont{J.~B.} \bibnamefont{Pendry}},
  \bibinfo{author}{\bibfnamefont{A.~F.} \bibnamefont{Starr}}, \bibnamefont{and}
  \bibinfo{author}{\bibfnamefont{D.~R.} \bibnamefont{Smith}},
  \bibinfo{journal}{Science} \textbf{\bibinfo{volume}{314}},
  \bibinfo{pages}{977} (\bibinfo{year}{2006}).

\bibitem[{\citenamefont{Cai et~al.}(2007)\citenamefont{Cai, Chettiar,
  Kildishev, and Shalaev}}]{Cai07}
\bibinfo{author}{\bibfnamefont{W.}~\bibnamefont{Cai}},
  \bibinfo{author}{\bibfnamefont{U.~K.} \bibnamefont{Chettiar}},
  \bibinfo{author}{\bibfnamefont{A.~V.} \bibnamefont{Kildishev}},
  \bibnamefont{and} \bibinfo{author}{\bibfnamefont{V.~M.}
  \bibnamefont{Shalaev}}, \bibinfo{journal}{Nat Photon}
  \textbf{\bibinfo{volume}{1}}, \bibinfo{pages}{224} (\bibinfo{year}{2007}).

\bibitem[{\citenamefont{Liu et~al.}(2008)\citenamefont{Liu, Guo, Fu, Kaiser,
  and Giessen}}]{Liu08}
\bibinfo{author}{\bibfnamefont{N.}~\bibnamefont{Liu}},
  \bibinfo{author}{\bibfnamefont{H.}~\bibnamefont{Guo}},
  \bibinfo{author}{\bibfnamefont{L.}~\bibnamefont{Fu}},
  \bibinfo{author}{\bibfnamefont{H.}~\bibnamefont{Kaiser},
  \bibfnamefont{Stefan~Schweizer}}, \bibnamefont{and}
  \bibinfo{author}{\bibfnamefont{H.}~\bibnamefont{Giessen}},
  \bibinfo{journal}{Nat Mater} \textbf{\bibinfo{volume}{7}},
  \bibinfo{pages}{31} (\bibinfo{year}{2008}).

\bibitem[{\citenamefont{Smolyaninov et~al.}(2008)\citenamefont{Smolyaninov,
  Hung, and Davis}}]{Smolyaninov08}
\bibinfo{author}{\bibfnamefont{I.~I.} \bibnamefont{Smolyaninov}},
  \bibinfo{author}{\bibfnamefont{Y.~J.} \bibnamefont{Hung}}, \bibnamefont{and}
  \bibinfo{author}{\bibfnamefont{C.~C.} \bibnamefont{Davis}},
  \bibinfo{journal}{Opt. Lett.} \textbf{\bibinfo{volume}{33}},
  \bibinfo{pages}{1342} (\bibinfo{year}{2008}),
  \urlprefix\url{http://ol.osa.org/abstract.cfm?URI=ol-33-12-1342}.

\bibitem[{\citenamefont{Smolyaninov et~al.}(2009)\citenamefont{Smolyaninov,
  Smolyaninova, Kildishev, and Shalaev}}]{Smolyaninov09}
\bibinfo{author}{\bibfnamefont{I.~I.} \bibnamefont{Smolyaninov}},
  \bibinfo{author}{\bibfnamefont{V.~N.} \bibnamefont{Smolyaninova}},
  \bibinfo{author}{\bibfnamefont{A.~V.} \bibnamefont{Kildishev}},
  \bibnamefont{and} \bibinfo{author}{\bibfnamefont{V.~M.}
  \bibnamefont{Shalaev}}, \bibinfo{journal}{Phys. Rev. Lett.}
  \textbf{\bibinfo{volume}{102}}, \bibinfo{pages}{213901}
  (\bibinfo{year}{2009}),
  \urlprefix\url{http://link.aps.org/doi/10.1103/PhysRevLett.102.213901}.

\bibitem[{\citenamefont{Landy and Smith}(2012)}]{Landy12}
\bibinfo{author}{\bibfnamefont{N.}~\bibnamefont{Landy}} \bibnamefont{and}
  \bibinfo{author}{\bibfnamefont{D.~R.} \bibnamefont{Smith}},
  \bibinfo{journal}{Nat Mater} \textbf{\bibinfo{volume}{advance online
  publication}}, \bibinfo{pages}{31} (\bibinfo{year}{2012}).

\bibitem[{\citenamefont{Liu et~al.}(2009)\citenamefont{Liu, Ji, Mock, Chin,
  Cui, and Smith}}]{Liu09}
\bibinfo{author}{\bibfnamefont{R.}~\bibnamefont{Liu}},
  \bibinfo{author}{\bibfnamefont{C.}~\bibnamefont{Ji}},
  \bibinfo{author}{\bibfnamefont{J.~J.} \bibnamefont{Mock}},
  \bibinfo{author}{\bibfnamefont{J.~Y.} \bibnamefont{Chin}},
  \bibinfo{author}{\bibfnamefont{T.~J.} \bibnamefont{Cui}}, \bibnamefont{and}
  \bibinfo{author}{\bibfnamefont{D.~R.} \bibnamefont{Smith}},
  \bibinfo{journal}{Science} \textbf{\bibinfo{volume}{323}},
  \bibinfo{pages}{366} (\bibinfo{year}{2009}).

\bibitem[{\citenamefont{Valentine et~al.}(2009)\citenamefont{Valentine, Li,
  Zentgraf, Bartal, and Zhang}}]{Valentine09}
\bibinfo{author}{\bibfnamefont{J.}~\bibnamefont{Valentine}},
  \bibinfo{author}{\bibfnamefont{J.}~\bibnamefont{Li}},
  \bibinfo{author}{\bibfnamefont{T.}~\bibnamefont{Zentgraf}},
  \bibinfo{author}{\bibfnamefont{G.}~\bibnamefont{Bartal}}, \bibnamefont{and}
  \bibinfo{author}{\bibfnamefont{X.}~\bibnamefont{Zhang}},
  \bibinfo{journal}{Nat Mater} \textbf{\bibinfo{volume}{8}},
  \bibinfo{pages}{568} (\bibinfo{year}{2009}).

\bibitem[{\citenamefont{Gabrielli et~al.}(2009)\citenamefont{Gabrielli,
  Cardenas, Poitras, and Lipson}}]{Gabrielli09}
\bibinfo{author}{\bibfnamefont{L.~H.} \bibnamefont{Gabrielli}},
  \bibinfo{author}{\bibfnamefont{J.}~\bibnamefont{Cardenas}},
  \bibinfo{author}{\bibfnamefont{C.~B.} \bibnamefont{Poitras}},
  \bibnamefont{and} \bibinfo{author}{\bibfnamefont{M.}~\bibnamefont{Lipson}},
  \bibinfo{journal}{Nat Photon} \textbf{\bibinfo{volume}{3}},
  \bibinfo{pages}{461} (\bibinfo{year}{2009}).

\bibitem[{\citenamefont{Leonhardt and Tyc}(2009)}]{Leonhardt09}
\bibinfo{author}{\bibfnamefont{U.}~\bibnamefont{Leonhardt}} \bibnamefont{and}
  \bibinfo{author}{\bibfnamefont{T.}~\bibnamefont{Tyc}},
  \bibinfo{journal}{Science} \textbf{\bibinfo{volume}{323}},
  \bibinfo{pages}{110} (\bibinfo{year}{2009}).

\bibitem[{\citenamefont{Zhang et~al.}(2011)\citenamefont{Zhang, Luo, Liu, and
  Barbastathis}}]{Zhang11}
\bibinfo{author}{\bibfnamefont{B.}~\bibnamefont{Zhang}},
  \bibinfo{author}{\bibfnamefont{Y.}~\bibnamefont{Luo}},
  \bibinfo{author}{\bibfnamefont{X.}~\bibnamefont{Liu}}, \bibnamefont{and}
  \bibinfo{author}{\bibfnamefont{G.}~\bibnamefont{Barbastathis}},
  \bibinfo{journal}{Phys. Rev. Lett.} \textbf{\bibinfo{volume}{106}},
  \bibinfo{pages}{033901} (\bibinfo{year}{2011}),
  \urlprefix\url{http://link.aps.org/doi/10.1103/PhysRevLett.106.033901}.

\bibitem[{\citenamefont{Fridman et~al.}(2012)\citenamefont{Fridman, Farsi,
  Okawachi, and Gaeta}}]{Fridman12}
\bibinfo{author}{\bibfnamefont{M.}~\bibnamefont{Fridman}},
  \bibinfo{author}{\bibfnamefont{A.}~\bibnamefont{Farsi}},
  \bibinfo{author}{\bibfnamefont{Y.}~\bibnamefont{Okawachi}}, \bibnamefont{and}
  \bibinfo{author}{\bibfnamefont{A.~L.} \bibnamefont{Gaeta}},
  \bibinfo{journal}{Nature} \textbf{\bibinfo{volume}{481}}, \bibinfo{pages}{62}
  (\bibinfo{year}{2012}).

\bibitem[{\citenamefont{Houdin}(2008)}]{Houdin1868}
\bibinfo{author}{\bibfnamefont{R.}~\bibnamefont{Houdin}},
  \emph{\bibinfo{title}{The secrets of stage conjuring}}
  (\bibinfo{publisher}{Wildside Press LLC}, \bibinfo{year}{2008}).

\bibitem[{Mir()}]{MirrorInvisibility}
\urlprefix\url{www.youtube.com/watch?v=_hF8xuPShsM,
  www.youtube.com/watch?v=RwgIr06OJLo}.

\bibitem[{\citenamefont{Wolf and Habashy}(1993)}]{Wolf93}
\bibinfo{author}{\bibfnamefont{E.}~\bibnamefont{Wolf}} \bibnamefont{and}
  \bibinfo{author}{\bibfnamefont{T.}~\bibnamefont{Habashy}},
  \bibinfo{journal}{Journal of Modern Optics} \textbf{\bibinfo{volume}{40}},
  \bibinfo{pages}{785} (\bibinfo{year}{1993}).

\end{thebibliography}

\end{document}